\documentclass[twocolumn,prd,groupedaddress,nofootinbib]{revtex4}

\usepackage{amsfonts,amsmath,amssymb,mathrsfs}
\usepackage{hyperref}
\usepackage{subfigure}
\usepackage{color}
\usepackage{graphicx}  % needed for figures
\usepackage{dcolumn}   % needed for some tables
\usepackage{bm}        % for math
\usepackage[english]{babel}

\def\a{\alpha}

\def\r{\rho}
\def\s{\sigma}
\def\t{\tau}
\def\m{\mu}
\def\n{\nu}
\def\k{\kappa}
\def\th{\theta}
\def\g{\gamma}\def\G{\Gamma}
\def\L{\Lambda}\def\l{V}
\def\D{\Delta}
\def\la{\langle}
\def\ra{\rangle}
\def\o{\omega}\def\O{\Omega}
\def\d{\delta}
\def\p{\partial}

\def\oxthree{{\cal O}(x^3) }

\def\half{\textstyle{\frac{1}{2}}}

\def\bdoc{\begin{document}}
\def\edoc{\end{document}}
\def\bea{\begin{equation}}
\def\eea{\end{equation}}

\def\beq{\begin{eqnarray}}
\def\eeq{\end{eqnarray}}
\def\ben{\begin{enumerate}}
\def\een{\end{enumerate}}
\def\la{\langle}
\def\ra{\rangle}
\def\a{\alpha}
\def\g{\gamma}\def\G{\Gamma}
\def\d{\delta}\def\D{\Delta}
\def\e{\epsilon}
\def\z{\zeta}

\def\th{\theta}
\def\k{\kappa}
\def\l{\lambda}
\def\m{\mu}
\def\n{\nu}
\def\o{\omega}
\def\p{\pi}
\def\r{\rho}
\def\s{\sigma}
\def\t{\tau}
\def\L{{\cal L}}
\def\S{\Sigma }
\def\gsim{\; \raisebox{-.8ex}{$\stackrel{\textstyle >}{\sim}$}\;}
\def\lsim{\; \raisebox{-.8ex}{$\stackrel{\textstyle <}{\sim}$}\;}
\def\gtrsim{\gsim}
\def\lessim{\lsim}
\def\loc{{\rm local}}
\def\vm{v_{\rm max}}
\def\bh{\bar{h}}
\def\del{\partial}
\def\nab{\nabla}
\def\half{{\textstyle{\frac{1}{2}}}}
\def\fourth{{\textstyle{\frac{1}{4}}}}

\def\bD{{\bf D}}
\def\bE{{\bf E}}
\def\bF{{\bf F}}
\def\bB{{\bf B}}
\def\bP{{\bf P}}
\def\bV{{\bf v}}
\def\bv{{\bf v}}
\def\bx{{\bf x}}
\def\by{{\bf y}}
\def\bz{{\bf z}}
\def\ba{{\bf a}}
\def\bd{{\bf d}}
\def\bs{{\bf s}}
\def\bn{{\bf n}}
\def\bp{{\bf p}}

\def\O{\Omega}

\def\br{{\bf r}}
\def\bnab{{\bf \nab}}

\def\tE{\tilde{E}}
\def\tL{\tilde{L}}
\def\Horava{Ho\v{r}ava }

\def\oxtwo{\mathscr{O}\left(x^2\right)}
\def\oxthree{\mathscr{O}\left(x^3\right)}
\def\oxfour{\mathscr{O}\left(x^4\right)}
\def\oxfive{\mathscr{O}\left(x^5\right)}

\def\ph{\phantom}

\begin{document}
\title{Physical process first law and increase of horizon entropy for black holes in Einstein-Gauss-Bonnet gravity }
\author{Ayan Chatterjee}\email{achatterjee@imsc.res.in} \author{Sudipta Sarkar}\email{sudiptas@imsc.res.in}
\affiliation{The Institute of Mathematical Sciences, Chennai, India}
\date{\today}
\begin{abstract}
We establish the physical process version of first law by studying small perturbations of a stationary black hole with regular bifurcation surface in Einstein-Gauss-Bonnet (EGB) gravity. Our result shows that when the stationary black hole is perturbed by a matter stress energy tensor and finally settles down to a new stationary state, the Wald entropy increases as long as the matter satisfies null energy condition. 
\end{abstract}
\maketitle

The striking similarity of the laws of black hole mechanics with thermodynamics was first established in case of general relativity (GR) \cite{Bardeen:1973gs}. It is interesting to explore whether this analogy is a peculiar property of GR or a robust feature of any generally covariant theory of gravity.\\
The zero'th law, which ensures the constancy of surface gravity is valid for any stationary Killing horizon with a regular bifurcation surface \cite{Racz:1992bp}, irrespective of the gravitational dynamics.\\
The equilibrium state version of first law is established by Wald and collaborators \cite{Wald:1993nt,Iyer:1994ys} for any arbitrary diffeomorphisom invariant theory of gravity. Comparing with the first law of thermodynamics, the entropy 
of the black hole can be expressed as a local geometric quantity integrated over a space-like cross section of the horizon and is associated with the Noether charge of Killing isometry that generates the horizon.\\
For black holes in GR, the analog of second law of thermodynamics is the ``area theorem'' which asserts that area of a black hole can not decrease in any classical process \cite{Hawking:1971vc}. A simple illustration of this result can be obtained by studying the physical process version of the first law which describes the dynamical change of horizon area in response to a flux of matter through the horizon \cite{Hawking:1972hy}. In case of stationary bifurcate horizons with a horizon generating Killing field $\xi^a$, the physical process first law is,
\beq
\Delta A = -\frac{8\pi}{\kappa}\int_{{\cal H}} T_{ab} \xi^a dH^b \label{physical_law}
\eeq
where $\kappa$ is the surface gravity of the black hole, $T_{ab}$ is the stress energy tensor of the matter and the integration is over the horizon ${\cal H}$. The  measure is $dH_b=-k_b \,d\lambda \sqrt{\gamma}\,dA$, where $\lambda$ is an affine parameter along the horizon generators, $k^b=(\partial_\lambda)^b$ is tangent to the generators, and $\sqrt{\gamma} dA$ is the area element of a constant $\lambda$ horizon slice.( For a detail discussion and derivation of this law for GR, see Refs. \cite{Jacobson:2003wv}, \cite{Amsel:2007mh} and \cite{Gao:2001ut}). The main ingredients in the derivation are the Raychaudhuri equation and the assumption of sufficiently quasi stationary process in which the expansion and shear of the generators are small and hence all second order terms can be neglected. There is also an assumption that the black hole is stable under small perturbations. Eq.(\ref{physical_law}) shows that in a dynamical process, the area of a black hole can not decrease as long as the matter satisfies null energy condition (i.e $T_{ab} k^a k^b \geq 0$). Once we identify the Hawking temperature as $T = \kappa/ 2 \pi$, Eq.(\ref{physical_law}) becomes completely analogous to the Clausius relationship $ \Delta S = \Delta Q / T$ as in ordinary thermodynamic systems.\\
 The question of validity of the full second law for arbitrary theory of gravity still remains an unresolved issue. Except for the case of $f(R)$-gravity \cite{Jacobson:1995uq}, there is no proof of the analog of Hawking's area theorem beyond GR. In the quasi-stationary
case, an argument for second law valid for all diffeomorphism invariant gravity theories was given in Ref. \cite{Jacobson:1995uq}, under the assumption that the stationary comparison version of the first law implies the physical process version for quasi-stationary processes. Still, we need a direct proof of the physical process version, by computing the change of the Wald entropy due to the accretion of matter by the black hole. This is an important consistency check for the validity of black hole thermodynamics beyond GR \cite{Wald:1995yp}. If the physical process and the equilibrium state versions of the first law do not match, this will suggest an inconsistency in the assumptions behind the derivation, in particular to the assumption that the black hole horizon is not destroyed by throwing matter into it and the final state in the late time is again stationary. The fact that in GR, both these versions agree, lends strong support to the idea that gravitational collapse result in a predictable black hole. It is expected that black holes in any reasonable theory of gravity must have such property. Also, a direct proof of physical process version of first law will automatically establish a quasi-stationary version of the second law \cite{Jacobson:1995uq}. \\
In this letter, we provide a direct proof of the physical process version of the  first law for Einstein-Gauss-Bonnet (EGB) gravity. Our proof involves the direct evaluation of the change of the  Wald entropy when a stationary black hole is weakly perturbed  and finally settles down to a new stationary state. We show that in such a process, the Wald entropy for EGB theory can not decrease as long as the matter stress energy tensor satisfies null energy condition. This firmly establishes that the Wald entropy for stationary black holes merits the name entropy even for more general gravity theories.\\
The letter is organized as follows: we begin by presenting the properties of stationary Killing horizons and discuss how Einstein tensor takes a highly symmetric form on the Killing horizon of any stationary (and non-extremal) black hole spacetime \cite{Medved:2004tp}. Next, we review the Wald entropy for stationary horizons in EGB theory and present the proof of the physical process version of the first law. Finally, we conclude with some discussions\footnote{We adopt the metric signature $(-, +, +, +, ...)$ and our sign conventions are same as those of \cite{Wald:1984rg}.}.\\
Let us begin with properties of a stationary black hole horizon. In a $D$-dimensional spacetime, the event horizon is a null hyper-surface ${\cal H}$ parametrized by an affine parameter $\lambda$. The vector field $k^a = (\partial_\lambda)^a$ is tangent to the horizon and obeys geodesic equation. Consider a spatial cross section at $\lambda = 0$ everywhere. All such $\lambda =$ constant slices are space-like and foliate the horizon. Any point $p$ on such slices have coordinates $\{\lambda, x^A\}$  where $x^A, \,(A=2, \cdots ,D)$ are the coordinates of a point on $\lambda = 0$ slice connected with $p$ by a horizon generator. We can construct a basis with the vector fields, $\{k^a, l^a, e^{a}_{A}\}$ where $l^a$ is a second null vector such that $l^a k_a = -1$. The induced metric on any slice is $\gamma_{ab} = g_{ab} + 2 k_{(a} l_{b)}$. The change of the induced metric from one slice to another can be obtained from the metric evolution equation \cite{Wald:1984rg}, 
\beq
{\cal L}_{k} \gamma_{ab} = 2 \left( \sigma_{ab} + \frac{\theta}{(D-2)} \gamma_{ab} \right),\label{metric_evolution}
\eeq
where $\sigma_{ab}$ is the shear and $\theta$ is the expansion of the horizon. If the event horizon is also a Killing horizon 
\footnote{ Here we make an implicit assumption, that the event horizon of a stationary black hole is also a Killing horizon with regular bifurcation surface. Although this is certainly true for GR \cite{Hawking:1973uf}, we are not aware of any proof for EGB gravity.}, i.e. the horizon generators are the orbits of a Killing field $\xi^a = (\partial/\partial v)^a$, which is null on the horizon, we can define the surface gravity $\kappa$ by the relationship $\xi^a \nabla_a \xi^b = \kappa\, \xi^b$. For stationary spacetimes with a Killing horizon, both the expansion and shear vanish and using Raychaudhuri equation and the evolution equation for shear, we obtain \cite{Wald:1984rg}, $R_{ab} k^a k^b = k^a k^c \gamma^{b}_{m} \gamma^{d}_{n} C_{abcd} = 0$ where $C_{abcd}$ is the Weyl tensor. \\
The Einstein tensor (and hence the Ricci tensor) takes a highly symmetric, block diagonal form on a stationary Killing horizon with a regular bifurcation surface \cite{Medved:2004tp}. To understand this result, let us construct a basis $\{\xi^a, N^a, e^{a}_A\}$, where $N^a$ is another null vector satisfying $\xi^a N_a = -1$ and express Ricci tensor in this basis. On the stationary Killing horizon, Ricci tensor satisfies the relation $R_{ab} \xi^b \sim \xi_a$ \cite{Wald:1984rg}. Using this, it is straightforward to show, that on the bifurcation surface, the Ricci tensor must have a form, $R_{ab} = C_1 g^{\perp}_{ab} + C_{ab}$, where $g^{\perp}_{ab} = -2 k_{(a} l_{b)}$, is the metric of the $2$-dimensional plane orthogonal to the $(D-2)$-dimensional bifurcation surface. The coefficient $C_1$ is given by, $C_1 = -R_{ab}k^a l^b$ and $C_{ab}$ is a tensor entirely intrinsic to the horizon cross section such that $k^a C_{ab} = l^a C_{ab} = 0$. Since Ricci tensor is a Killing invariant, we can Lie propagate the Ricci tensor onto any other space-like section of the horizon (For a related discussion, see \cite{Jacobson:1993vj} ). For stationary Killing horizons, since both the expansion and shear vanish, using the block diagonal form of the Ricci tensor on the horizon, it is also possible to show that \cite{Ortaggio:2007eg}, $k^a \gamma^{b}_{i}\gamma^{c}_{j} \gamma^{d}_{k} C_{abcd} = 0$.\\
Another relationship which will be useful for our calculation is the expression of intrinsic Ricci curvature $^{(D-2)}R^{ab}$ of the cross-section of the
stationary Killing horizon in terms of full curvatures, given by  \cite{Gourgoulhon:2005ng}\,
\beq
^{(D-2)}R_{ab} = \gamma^{m}_{a}\, \gamma^{n}_{b}\,\gamma^{r}_{s} \, R^{s}_{mrn}. \label{curvature_decomposition}
\eeq
Next, we discuss the features of EGB gravity theory. A simple modification of the Einstein-Hilbert action is to include the  higher order curvature terms preserving the diffeomorphism invariance and still leading to an equation of motion containing no more than second order time derivatives. In fact this generalization is unique \cite{Lovelock:1971yv} and the lowest order correction appears as the Gauss-Bonnet (GB) term, ${\cal L}_{GB} = R^2 - 4 R_{ab} R^{ab} + R_{abcd}R^{abcd}$ in spacetime dimensions $D>4$. EGB gravity is free from ghosts \cite{Zwiebach:1985uq, Zumino:1985dp}, leads to a well-defined initial value problem and therefore,  is a reasonable candidate for a low energy effective theory of gravity. The action functional is given by,
\beq
{\cal L} = \frac{1}{16 \pi} \int d^D x \sqrt{-g} ~\left( R  + \alpha {\cal L}_{GB} \right) \label{actionGB}.
\eeq
The field equation of EGB theory is, $G_{ab} + \alpha H_{ab} = 8 \pi T_{ab}$ where,
\begin{eqnarray}
{H}_{ab}&\equiv&2\Bigl[RR_{ab}-2R_{aj}R^j_{~b}
-2R^{ij}R_{aibj}
\nonumber
\\
&& ~~~~
 +R_{a}^{~ijk}R_{bijk}\Bigr]
-\frac{1}{2}g_{ab} {\cal L}_{GB}.
\end{eqnarray}
For the action in Eq.(\ref{actionGB}), the Wald entropy associated with a stationary Killing horizon is \cite{Jacobson:1993xs} ,
\beq
S = \frac{1}{4} \int \rho~ \sqrt{\gamma}~dA \label{entropyGB},
\eeq
where the entropy density $\rho = \left( 1 + 2 \alpha ~^{(D-2)}R \right)$ and the integration is over $(D-2)$-dimensional space-like cross-section of the horizon. $^{(D-2)}R$ is the intrinsic Ricci scalar of the horizon cross-section. \\
Our goal is to prove that this entropy always increases when a black hole is perturbed by a weak matter stress energy tensor of order ${\cal O}(\epsilon)$ provided the matter obeys null energy condition. Since the black hole is stationary in the asymptotic future, the vector field $\xi^a$ is an exact Killing vector at late times and all the Lie derivatives of dynamical fields w.r.t $\xi^a$ vanish. The accretion process is assumed to be slow such that all changes of the dynamical fields are first order and we can neglect all viscous effects. More specifically,
 we assume that:
\beq
 \theta \sim \sigma_{ab}\sim {\cal O}(\epsilon).
\eeq
 The bifurcation surface at $\lambda = 0$ is taken as the initial cross-section and the final cross-section is in the asymptotic stationary regime. Then the change is entropy is \cite{Jacobson:1995uq} ,
\beq\label{change_wald_ent}
\Delta\,S &=&\frac{1}{4}\int_{{\cal H}} \left(\frac{d\rho}{d\lambda} +\theta\, \rho \right)\,d\lambda \,\sqrt{\gamma}\,
 dA, \nonumber\\
 &\simeq &-\frac{1}{4}\int_{{\cal H}}  \left(\frac{d^{2}\rho}{d\lambda^{2}} - \rho~ R_{ab}k^{a} k^{b}\right)\,\lambda\, d\lambda \,\sqrt{\gamma}\,dA
\eeq 
In deriving the second line, a total derivative is discarded since
it vanishes both on the initial $\lambda=0$ slice and on the final
stationary state. Also, we have  used the Raychaudhuri equation and
 neglected all terms except those first order in perturbation. Using field equation, this entropy change is expressed as,
\beq
\Delta S &=& \int_{{\cal H}} \lambda \left (2 \pi \, T_{ab}
-\frac{\alpha}{4}{\cal R}_{ab}\right)~ k^{a}k^{b}d\lambda
\,\sqrt{\gamma}\,dA, \label{entropy_change}
\eeq
where ${\cal R}_{ab} = H_{ab} - 2 ^{(D-2)}R R_{ab} + 2\,\nabla_{a}\nabla_{b}\, ^{(D-2)}R$. The first term in Eq.(\ref{entropy_change}) is linear in perturbation. Our aim is to prove that the first order part of the term ${\cal R}_{ab} k^a k^b$ vanishes identically. In order to see that, we note that the terms in ${\cal R}_{ab}$ involve squares of curvatures. We are only interested in quantities first order in perturbation over a background stationary spacetime. The background spacetime in our context has no particular physical meaning except as a reference spacetime. Therefore, when we encounter a product of two quantities $X$ and $Y$, to extract the part linear in perturbation, we will always express such a product as,
\beq 
X Y \approx  X^{(B)} \, Y^{(P)} +  X^{(P)} \,Y ^{(B)},\label{perturbation scheme} 
\eeq
where $X^{(B)}$ is the value of the quantity $X$ evaluated on the stationary background and $X^{(P)} $ is the perturbed value of $X$ linear in perturbation. Note that, on the stationary background, Raychaudhuri equation demands $R^{(B)}_{ab} k^a k^b = 0$ and since $T^{(B)}_{ab} k^a k^b = 0$, we have $H^{(B)}_{ab} k^a k^b = 0$. Also, to simplify the calculation, we use diffeomorphism freedom to make the null geodesic generators of the event horizon of the perturbed black hole coincide with the null geodesic generators of the background stationary black hole \cite{Gao:2001ut}. As a result, the perturbation in the location of the horizon vanishes.\\
As an illustration of the perturbation scheme mentioned in the Eq.(\ref{perturbation scheme}), we evaluate a term in $H_{ab} k^a k^b$ as,
\beq
&&2 R^{ij}R_{aibj} k^{a}k^{b}= 2\left(R^{(B)ij}R^{(P)}_{aibj}+
R^{(P)ij}R^{(B)}_{aibj}\right)\,k^{a}k^{b}\nonumber\\
&&~~~~~~~~~~~~=2 \left(C^{mn} R^{(P)}_{ambn}
+R^{(P)ij}\,R^{(B)}_{aibj}\right)\,k^{a}k^{b},
\eeq
where we have used the results of \cite{Medved:2004tp} to express the background Ricci tensor on the horizon as $R^{(B)}_{ab} = C_1 g^{\perp}_{ab} + C_{ab}$. Also, note that $\gamma^{ij}R^{(P)}_{aibj} k^a k^b =  R^{(P)}_{ab} k^a k^b$. Implementing similar scheme for other terms, the first order part ${\cal R}_1 $ of $\left( H_{ab} - 2 ^{(D-2)}R R_{ab} \right) k^a k^b$ becomes,
\beq
{\cal R}_1  &=& \left[-\frac{ 8\,R^{(B)}}{(D-1)(D-2)} + \frac{ 8\,C_1 }{(D-2)}  \right. \nonumber \\ &{}&~~~~~~ -\left. \frac{8 \,C^{(B)}_{abcd}\, k^a l^b l^c k^d}{(D-2)}\right] R^{(P)}_{mn}\,k^m k^n \nonumber \\ &&~~~~~~~~~-\frac{4(D-4)}{(D-2)} C^{ab} R^{(P)}_{aibj} k^i k^j. \label{1stterm}
\eeq
The last term in ${\cal R}_{ab} k^a k^b$ requires special care. To evaluate this term, we note that the change of the $(D-2)$-dimensional scalar curvature can be thought of due to the change in the intrinsic metric. Then, we can calculate this change by using the standard result of variation of Ricci scalar as,
\beq
k^a \nabla_a ( ^{(D-2)}R) &=& \frac{d ^{(D-2)}R}{d\lambda} \nonumber \\ &=& ^{(D-2)}R_{ab}\, {\cal L}_{k} \gamma^{ab} + D_a (\delta_{\lambda} V^a), \label{variation_section}
\eeq
\\
where $D_{a}$ is the covariant derivative intrinsic to the horizon cross-section and since the sections of the horizon are compact surfaces without boundaries, the surface term $D_a (\delta_{\lambda} V^a)$ in Eq.(\ref{variation_section}) does not contribute. Then using Eq.(\ref{metric_evolution}) and neglecting terms of higher order, we obtain,
\beq
\frac{d^2 }{d\lambda^2} \left(^{(D-2)}R \right) &=& -2 ^{(D-2)}R^{(B)ab} \left(\frac{d\sigma_{ab}}{d\lambda} + \frac{\gamma_{ab}}{(D-2)}\frac{d \theta}{d\lambda} \right)\nonumber \\
&=& 2\,^{(D-2)}R^{(B)ab} \,
R^{(P)}_{acbd} k^{c}k^{d}, \label{3rdterm}
\eeq     
where we have again used Raychaudhuri equation and the evolution equation for shear \cite{Wald:1984rg} keeping terms linear in perturbation. The last line follows after expressing the perturbed Weyl tensor in terms of curvature and Ricci tensors.\\
We rewrite  Eq.(\ref{3rdterm}) using Eq.(\ref{curvature_decomposition}) and express $R^{(B)}_{smrn}$ in terms of Weyl and Ricci tensors of the background to arrive at,
\beq
\frac{d^2 }{d\lambda^2} \left(^{(D-2)}R \right) &=& \left[\frac{4\,C^{(B)}_{abcd}\, k^a l^b l^c k^d }{(D-2)} + \frac{4\, C}{(D-1)(D-2)} \nonumber \right. \\  &{}& -\left. \frac{4 \, C_1\,(D-3)}{(D-1)(D-2)}\right]R^{(P)}_{mn}\,k^m k^n \nonumber \\  && + \frac{2(D-4)}{(D-2)} C^{ab} R^{(P)}_{aibj} k^i k^j,
\eeq
where we have defined, $C = \gamma^{ab} C_{ab}$. Next, using the expression, $ R^{(B)} = 2 C_1 + C $, and comparing with Eq.(\ref{1stterm}), it is straightforward to prove that the first order term in ${\cal R}_{ab} k^a k^b$ vanishes identically and we have the result,
\beq
\Delta S &=& 2 \pi \int_{{\cal H}} \lambda\, T_{ab} k^{a}k^{b}\,d\lambda
\,\sqrt{\gamma}\,dA + {\cal O}(\epsilon^2). \label{final_form}
\eeq
For the background stationary horizon, the Killing vector $\xi^a$ is related with the horizon generators $k^a$ as $ \xi^a = \kappa \lambda k^a$, since $T_{ab}$ itself is of ${\cal O}(\epsilon)$, we can use this relation in  Eq.(\ref{final_form}). Then, identifying the Hawking temperature as $T = \kappa / 2 \pi$, in the leading order of perturbation, we finally obtain,
\beq
T \Delta S = - \int_{{\cal H}}  T_{ab}\, \xi^{a} \,dH^b.
\eeq
This is the desired form which establishes the physical process version of the first law for EGB gravity. If the matter stress tensor obeys null energy condition, Eq.(\ref{final_form}) shows that the Wald entropy for stationary black holes in EGB theory can not decrease in a dynamical process which perturbs the black hole and leads to a new stationary state. Since, it is expected that classical matter obeys null energy condition, we can conclude that as in case of GR, the entropy of stationary black holes in EGB theory can not decrease through any classical process.\\
An interesting feature of this derivation of physical process first law is that it is completely local in nature. Unlike the equilibrium state version, there is no reference of the asymptotic infinity. In fact, it is the field equation which enforces the first law for quasi stationary changes of the horizon. In case of GR, it is possible to reverse this argument and derive Einstein equation as an equation of state of the spacetime \cite{Jacobson:1995ab}. Attempting similar construction for EGB gravity will require a notion of entropy for non-stationary states. For some possible proposals, see Refs. \cite{Padmanabhan:2009vy} and \cite{Guedens:2011dy}.\\
The derivation of the laws of black hole mechanics is entirely classical. To complete the thermodynamic analogy, we need to invoke quantum theory. The idea that black holes  radiate at a temperature $ T = \kappa / 2\pi$ is entirely a consequence of quantum field theory in the presence of a horizon and independent of the gravitational dynamics. As a result, the mere analogy of classical black hole physics with thermodynamics becomes an exact law once the quantum effects are taken into consideration. Then, a natural interpretation of the black hole entropy is that it counts the quantum micro-states of the black hole. For any reasonable theory of gravity, which has stable black holes, the density of states must be of the form $\exp{(S_{W})}$, where $S_{W}$ is the Wald entropy. In fact, in the context of string theory, it has been shown at least for extremal and near-extremal black holes in EGB theory, that the microscopic computations exactly matches with the Wald formula \cite{Sen:2007qy,Lopes Cardoso:1998wt}. Also, the GB term appears as a low energy $\alpha'$ correction in case of tree level heterotic string theory \cite{Zwiebach:1985uq}. Hence, from this microscopic point of view, it is desirable that the Wald entropy of the black holes in EGB theory should increase and our result establishes this at least in the context of first order perturbation.\\
For ordinary thermodynamic systems, the entropy, by construction is a path independent function and the change of the entropy from one stationary state to another is always given by the difference of entropies between two states, independent of the process. The agreement of the equilibrium state and the physical process version of first law establishes the same property for black hole entropy. Hence, our result implies that the thermodynamic analogy for black holes is not only a feature of GR and simple $f(R)$ theories, but equally applicable for well-motivated modifications of GR, like the EGB theory. For $f(R)$ gravity, it is not very surprising that the physical process version holds, since all $f(R)$ theories are related to GR by a field dependent conformal rescaling of the metric. But, the validity of physical process version of first law for EGB gravity is a non-trivial check suggesting a more general nature of black hole thermodynamics.\\
A possible generalization of our work would be to study black holes in a general Lovelock theory \cite{Lovelock:1971yv}. For any $m$-th order Lovelock term ${\cal L}_m$ in $D$-spacetime dimensions, the entropy density of the stationary black holes is \cite{Jacobson:1993xs} simply proportional to $^{(D-2)}{\cal L}_{(m -1)}$, the previous Lovelock term but intrinsic to the horizon cross-section. This remarkable property suggests that our 
method can be readily used to investigate the physical process version of the first law for any Lovelock theory. \\
In this letter, we have only considered the increase of the black hole entropy alone, neglecting the contribution of the quantum fields outside the horizon. In GR, we know that the effective negative energy flux from quantum fields to a black hole can lead to a decrease of horizon entropy. Then the relevant question is whether a generalized second law $(\Delta(S_{BH} + S_{\textrm{outside}} ) \geq 0)$ holds. There are arguments \cite{Unruh:1982ic} that the
generalized Second Law applies for semi-classical processes in case of black holes in GR. Validity of these arguments for reasonable higher curvature gravity theories is still an open problem.\\ 
Finally, it is worthwhile to study whether the physical process  first law holds for any diffeomorphism invariant gravity theory or applies to a special class of action functional. The answer, in either way, will be an important input for any quantum theory of gravity.
\acknowledgments
We would like to thank G. Date, Raf Guedens, Sanved Kolekar, Ted Jacobson and T Padmanabhan for comments and discussions.

\end{document}